\documentclass[prd,aps,showpacs,twocolumn,nofootinbib]{revtex4}

\usepackage{graphics}
 
\newcommand{\be}{\begin{equation}}
\newcommand{\ee}{\end{equation}}
\newcommand{\bea}{\begin{eqnarray}}
\newcommand{\eea}{\end{eqnarray}}
\newcommand{\ba}{\begin{array}}
\newcommand{\ea}{\end{array}}

\newcommand{\fpi}{f_{\pi}}

\newcommand{\cotWT}{\cot\omega_{\rm WT}}

\begin{document}

\title{WChPT analysis of twisted mass lattice data}

\author{Sinya Aoki$^{1,2}$ and Oliver B\"ar$^{3}$}
\thanks{\emph{}Talk given by O.\ B\"ar at QNP '06, Madrid.}

\affiliation{
$^1$Graduate School of Pure and Applied Sciences, University of Tsukuba,
Tsukuba 305-8571, Ibaraki Japan \\
$^2$ Riken BNL Research Center, Brookhaven National Laboratory, Upton,
NY 11973, USA \\
$^3$Institute of Physics, Humboldt University Berlin, Newtonstrasse
15, 12489 Berlin, Germany
}

\date{October 2006}

\begin{abstract}
We perform a Wilson Chiral Perturbation Theory (WChPT) analysis of quenched twisted mass lattice data. The data were generated by two independent groups with three different choices for the critical mass. For one choice, the so-called pion mass definition, one observes a strong curvature for small quark masses in various mesonic observables (''bending phenomenon''). Performing a combined fit to the next-to-leading (NLO) expressions, we find that WChPT describes the data very well and the fits provide very reasonable values for the low-energy parameters. 
\end{abstract}
 
 
\maketitle

\section{Introduction}
\label{intro}

Twisted mass lattice QCD  \cite{Frezzotti:2001ea,Frezzotti:2000nk} has many advantages for numerical lattice simulations, with automatic O($a$) improvement at maximal twist \cite{Frezzotti:2003ni,Frezzotti:2004wz,Frezzotti:2005zm} being probably the most striking one (for a recent review see Ref.\ \cite{Shindler:2005vj}).
Maximal twist is achieved by tuning the bare untwisted mass $m_{0}$ to a particular (critical) value such that some matrix element vanishes. The condition for maximal twist is not unique and various choice have been employed in quenched simulations \cite{Bietenholz:2004wv,Abdel-Rehim:2005gz,Jansen:2005gf,Jansen:2005kk,Abdel-Rehim:2006ve}.

A puzzle observed in early quenched simulations is the so-called ``bending phenomenon''  \cite{Bietenholz:2004wv}.
Employing the pion mass definition for maximal twist (i.e.\  tuning $m_{0}$ to the value where the pion mass would vanish without a twisted mass term), one observed a strong curvature in the quark mass dependence in many observables ($m_{\pi},\fpi,m_{\rho}$) for small twisted quark masses $\mu$. This unexpected observation spurred further numerical simulations with other definitions for maximal twist  \cite{Abdel-Rehim:2005gz,Jansen:2005gf,Jansen:2005kk} as well as theoretical studies. Nowadays the bending phenomenon seems well understood both in terms of the Symanzik effective theory \cite{Frezzotti:2005gi} as well as in Wilson Chiral Perturbation theory (WChPT) \cite{Aoki:2004ta,Sharpe:2005rq,Aoki:2006nv} (for introductions to lattice ChPT see also Refs.\ \cite{Bar:2004xp,Sharpe:2006pu}).

It is a pleasant side effect of this effort to understand the bending phenomenon that there are lots of data available for five different lattice spacings and three definitions of maximal twist. 
Moreover, light quark masses could be reached with values for  $m_{\pi}/m_{\rho}$ down to $\sim 0.3$, where one might expect WChPT to provide an effective description of the data. In particular,
the characteristic curvature of the bending phenomenon provides a distinctive test for WChPT to pass if it is the correct low energy effective theory for twisted mass lattice QCD. 
Provided WChPT describes the data very well we may also obtain estimates for some combinations of low energy constants of the effective theory. This was sufficient motivation for us to carry out a WChPT analysis of the existing lattice data. Preliminary results involving data for two definitions of maximal twist only can be found in Ref.\ \cite{Aoki:2005ii}.

%===================
\section{Fitting the data}
\label{sec:2}
%===================
We analyzed quenched lattice data  generated by two different groups
\cite{Bietenholz:2004wv,Abdel-Rehim:2005gz,Jansen:2005gf,Jansen:2005kk} 
with the Wilson plaquette action at $\beta =5.85$ ($a\approx 0.123$fm) and $\beta= 6.0$ ($a\approx 0.093$fm). The standard Wilson fermion action with a twisted mass term was employed to calculate a variety of mesonic observables. The twisted quark mass covers the range
$m_{\pi}\approx 270 \ldots 1200 \,{\rm MeV} $, or, alternatively, 
$m_{\pi}/m_{\rho} \approx 0.3 -0.8$. Besides the pion mass there is data available for the pseudoscalar decay constant, the angle $\cotWT$, as well as some more observables which we did not analyze. 
The untwisted bare quark mass was tuned according to three different definitions of maximal twist:
the pion mass definition, the PCAC mass definition and the parity conservation definition. 
In total there are at most 52 data points available for each lattice spacing.

%=============
%: figure 1
%=============
\begin{figure*}[t]
\begin{center}
\resizebox{0.92\textwidth}{!}{
 \includegraphics{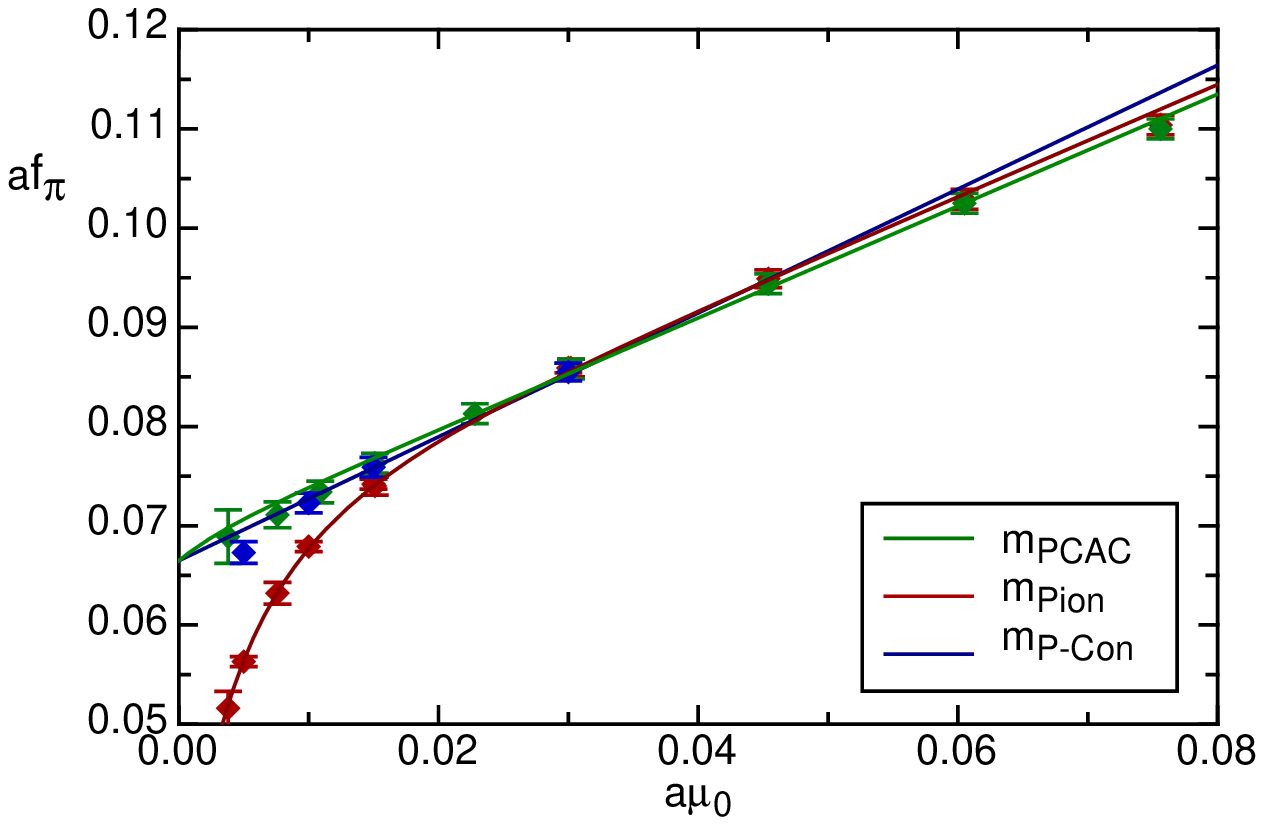} \hspace{1.9cm}  \includegraphics{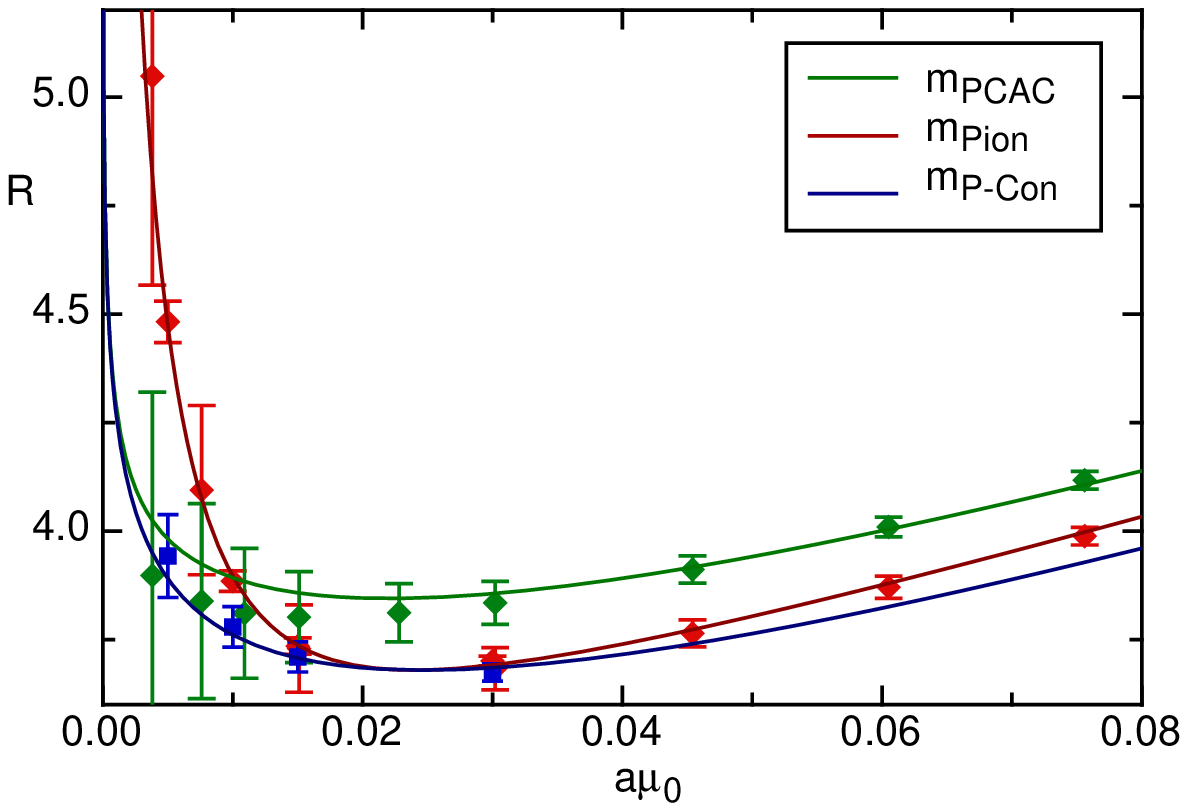}
 }
 \resizebox{0.92\textwidth}{!}{
  \includegraphics{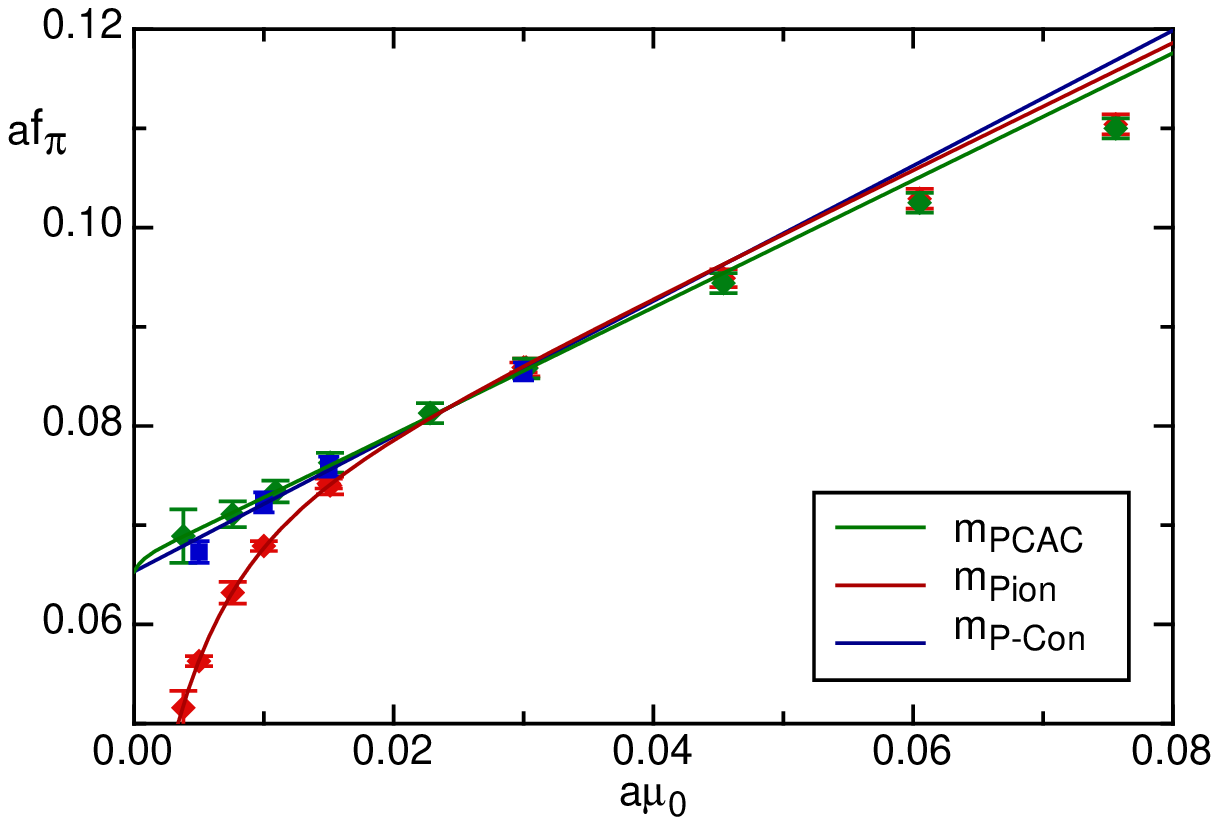}  \hspace{1.9cm}  \includegraphics{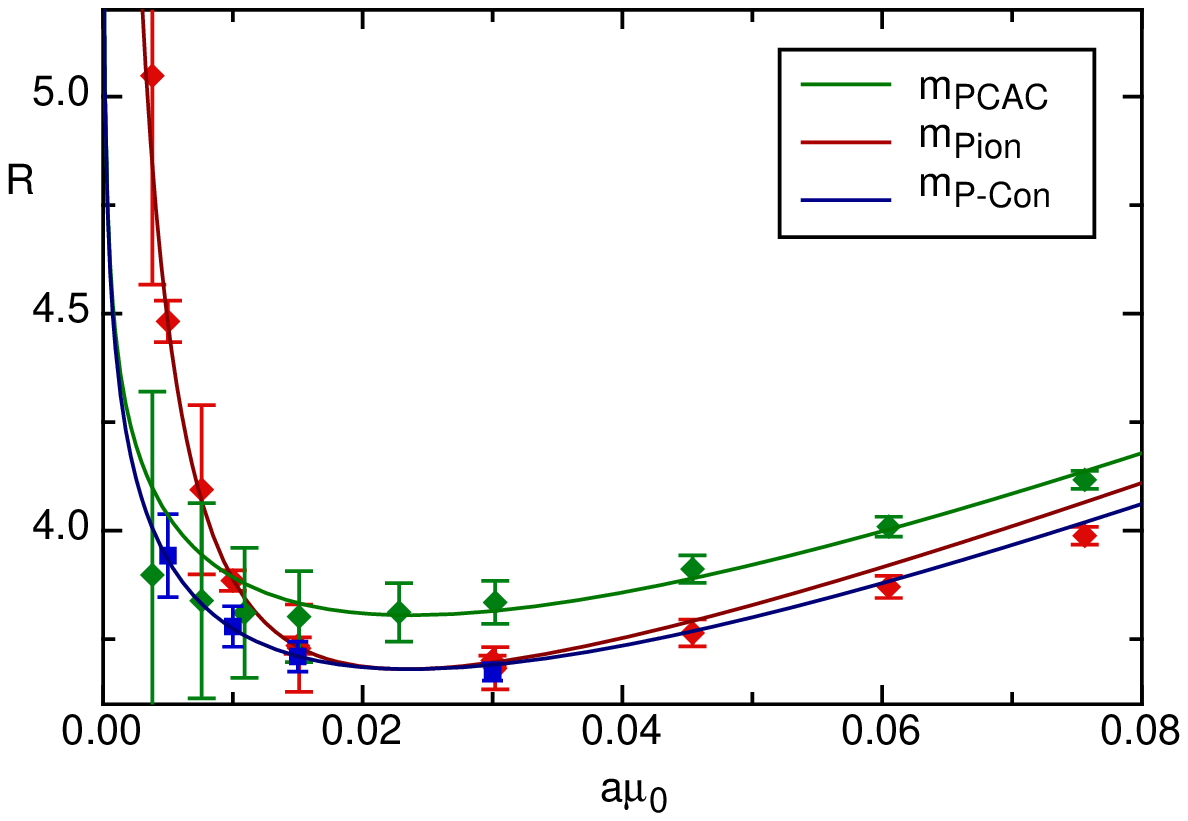}
}
\end{center}
\vspace*{0.3cm}   
\caption{Results of a combined fit for $\fpi$ and $R$ at $\beta=6.0$ ($\chi^{2}/{\rm d.o.f.} = 0.23$ with ${\rm d.o.f.} = 26$). The upper plots show the results where all data points are included in the fit, while data points with $a\mu_{0}\leq 0.0302$ only are included in the fits shown in the lower plots. 
}
\label{fig:1}      
\end{figure*}
%=============

There exists a vast literature  on WChPT for twisted mass lattice QCD \cite{Munster:2003ba,Scorzato:2004da,Aoki:2004ta,Sharpe:2004ny,Sharpe:2005rq}, which contains expressions for various mesonic observables up to next-to-leading order (NLO). The lattice artifacts are included through order O($a^{2}$) for different power countings and definitions of maximal twist. Here we use the NLO formulae given in Ref.\ \cite{Aoki:2006nv}, which include all NLO terms consistently for the two regimes where either $\mu\sim a$ or  $\mu\sim a^{2}$. The quenched chiral logarithm \cite{Bernard:1992mk,Sharpe:1992ft} is also included in the formulae. 

We performed combined fits of the WChPT formulae at NLO to the data for three observables: $\fpi$,
\bea
R & = & \frac{(a\,m_{\pi})^{2}}{a\mu}\,,\\
\cotWT & = & \frac{\langle \partial_{\mu}A_{\mu}^{1}P^{1}\rangle}{ \langle \partial_{\mu}V_{\mu}^{2}P^{1}\rangle}\,,
\eea
where $V^{a}_{\mu}$ and $A^{a}_{\mu}$ denote the (nonsinglet) vector and axial vector current, respectively.
At NLO we have in total  thirteen free fit parameters. Even though this is a fairly large number it is still small compared to the number of data points. 

We performed various fits, starting with all data points included and then successively remove the data points at high quark masses. In all cases we obtain good fit results with $\chi^{2}/{\rm d.o.f.}\approx0.2 - 0.5$, even if all data points up to $m_{\pi}/m_{\rho} \approx 0.8$ are included.\footnote{Note that the $\chi^{2}$ value is underestimated since the data is highly correlated.} 
Since we do not trust ChPT to work at such high masses we prefer to drop the data for the highest three masses. The fit results for $\fpi$ and $R$ at $\beta =6.0$ are shown in figure \ref{fig:1}. 
Even in this fit the heaviest point corresponds to $m_{\pi}/m_{\rho} \approx 0.63$, which is still heavy. Dropping more data points, however, makes the fit more and more unstable, so we cannot reduce the number of data points much further.

Apparently WChPT describes the data very well. In particular the bending for small masses in case of the pion mass definition is very well reproduced. This feature is independent of the number of data points included in the fit, even though the values for the fit parameters are different (see below).  
Note that the curvature in the data for $R$ with $a\mu_{0}\geq 0.3$ is also well described even though the heavier data points are excluded from the fit. 

The fits give reasonable values for the fit parameters. For the quenched chiral log parameter $\delta_{0}$, for example, we find
\bea
\delta_{0}& = & \left\{
\begin{array}{lclcl}
0.10\pm 0.03&\quad & \beta & = & 6.0\\[1ex]
0.054\pm 0.011&\quad & \beta & = & 5.85\\
\end{array}\right.
\eea
which is in very good agreement with the results obtained by other groups (for a summary see Ref.\ \cite{Wittig:2002ux}). 

We also obtain an estimate for the low-energy constant $c_{2}$. This parameter was first introduced in Ref.\ \cite{Sharpe:1998xm} and enters the chiral Lagrangian according to\footnote{The definition of $c_{2}$ is not unique and it is sometimes defined differently, for example in Ref.\ \cite{Aoki:2004ta}.} 
\bea
L_{\chi} & = & \ldots + \frac{c_{2}}{16}  \left\{\mbox{tr}(\Sigma + \Sigma^{\dagger})\right\}^{2} \ldots\,.
\eea
The sign of $c_{2}$ determines the phase diagram of the lattice theory and the pion mass splitting $\Delta m_{\pi}^{2}=m_{\pi^{0}}^{2} - m_{\pi^{\pm}}^{2}$ in the chiral limit is given by $c_{2}/f_{\pi}^{2}$ \cite{Sharpe:1998xm}.

From our fits we obtain for $c_{2}$ the value
\bea\label{eq:c2}
c_{2}& = & 
\begin{array}{lclcr}
[291 {\rm MeV}^{+4\%}_{-5\%}]^{4}&\quad & \beta & = & 5.85
\end{array}
\eea
This is the first determination of this low-energy constant, so we cannot compare with other results. 
However, the value seems reasonable based on dimensional analysis arguments. 
The fit for the smaller lattice spacing with $\beta = 6.0$ does not  determine $c_{2}$ very well; for the mean value we obtain approximately  $[170$ MeV$]^{4}$, but the error is of the same size. 

The physical parameters, on the other hand, are very well determined by the fit. For the pseudoscalar decay constant in the chiral limit we find
\bea
f_{0}& = & \left\{
\begin{array}{lclcl}
141.2 {\rm MeV}\pm 1\%&\quad & \beta & = & 6.0\\[1.6ex]
141.4 {\rm MeV}\pm 1\%&\quad & \beta & = & 5.85
\end{array}\right.
\eea
which is in very good agreement with earlier determinations. 
For the low-energy constant $\alpha_{5}^{\rm q}$ \cite{Heitger:2000ay}, entering the NLO expression for the decay constant, we obtain
\bea
\alpha_{5}^{\rm q}& = & \left\{
\begin{array}{lclcl}
1.03(5)&\quad & \beta & = & 6.0\\[1.6ex]
0.97(4) &\quad & \beta & = & 5.85
\end{array}\right.
\eea
Also these values agree very well with previous results in Ref.\ \cite{Heitger:2000ay}. 
Note that the results for $f_{0}$ and $\alpha_{5}^{\rm q}$ do not show any significant dependence on the lattice spacing. This is expected if WChPT works, since the main dependence on $a$ is captured by other terms in the chiral expansion, being directly proportional to powers in $a$.

We emphasize that the errors we quoted so far are only the statistical errors given by MINUIT which we used to perform the fits. These errors are underestimated due to the highly correlated 
data and the true statistical error can be substantially larger.
A second error source are systematic uncertainties, induced, for example, by the number of data points included in the fit. It is not simple to give a precise estimate for this error but we observed that the central value for $f_{0}$ changes by roughly  3 percent for the different fits we performed, while $\alpha_{5}^{\rm q}$ varies by about 7 percent.

%===================
\section{Conclusions}
\label{concl}
%===================
We performed fits of WChPT to quenched twisted mass data for $m_{\pi}^{2}, \fpi$ and the Ward-Takahashi angle $\cotWT$. We find that the NLO expressions describe the data very well with small $\chi^{2}$ values and reasonable values for the low-energy parameters. In particular, the bending phenomenon in case of the pion mass definition is very well reproduced.

The bending phenomenon is a very characteristic feature of twisted mass lattice QCD.  
It is encouraging that WChPT describes this distinct curvature very well. This indicates that WChPT, i.e.\ ChPT for lattice QCD, seems to work. Previous studies \cite{Farchioni:2004tv,Aoki:2003yv,Namekawa:2004bi}, using untwisted Wilson fermions, came to contradicting results and were not conclusive at all.  

So far we performed separate fits for each lattice spacing. In a next step it would be very interesting to perform a combined fit to the entire data set and take the continuum limit. These results should be compared to the results one obtains after a standard continuum extrapolation where one assumes a polynomial lattice spacing dependence. This would partly answer the question whether WChPT is not only able to describe the lattice data but also necessary to extract the correct continuum physics from the data.

\end{document}